\documentclass[twocolumn]{aastex62}

\usepackage[para]{threeparttable}
\usepackage{comment}
\usepackage{tabularx}
\usepackage{color}

\newcommand{\Suzaku}{{\it Suzaku}}

\newcommand{\XMM}{{\it XMM--Newton}}
\newcommand{\NH}{$N_{\rm H}$}

\graphicspath{{./}{figures/}}

\received{April 2, 2021}
\revised{May 5, 2021}
\accepted{May 8, 2021}
\submitjournal{ApJL}

\shorttitle{Highly-Neutronized Ejecta in 3C\,397}
\shortauthors{Ohshiro et al.}

\begin{document}

\title{Discovery of the Highly-Neutronized Ejecta Clump with Enhanced Abundances of 
Titanium and Chromium in the Type I\lowercase{a} Supernova Remnant 3C\,397}

\correspondingauthor{Yuken Ohshiro}
\email{ohshiro-yuken@g.ecc.u-tokyo.ac.jp}

\author{Yuken Ohshiro}
\affiliation{Department of Physics, Graduate School of Science, The University of Tokyo, 7-3-1 Hongo, Bunkyo-ku, Tokyo 113-0033, Japan}
\affiliation{Institute of Space and Astronautical Science (ISAS), Japan Aerospace Exploration Agency (JAXA), 3-1-1 Yoshinodai, Chuo-ku, Sagamihara, Kanagawa 252-5210, Japan}

\author{Hiroya Yamaguchi}
\affiliation{Institute of Space and Astronautical Science (ISAS), Japan Aerospace Exploration Agency (JAXA), 3-1-1 Yoshinodai, Chuo-ku, Sagamihara, Kanagawa 252-5210, Japan}
\affiliation{Department of Physics, Graduate School of Science, The University of Tokyo, 7-3-1 Hongo, Bunkyo-ku, Tokyo 113-0033, Japan}

\author{Shing-Chi Leung}
\affiliation{TAPIR, Walter Burke Institute for Theoretical Physics, Mailcode 350-17, Caltech, Pasadena, CA 91125, USA}

\author{Ken’ichi Nomoto}
\affiliation{Kavli Institute for the Physics and Mathematics of the Universe (WPI), The University of Tokyo, Kashiwa, Chiba 277-8583, Japan}

\author{Toshiki Sato}
\affiliation{Department of Physics, Rikkyo University, 3-34-1 Nishi Ikebukuro, Toshima-ku, Tokyo 171-8501, Japan}

\author{Takaaki Tanaka}
\affiliation{Department of Physics, Konan University, 8-9-1 Okamoto, Higashinada, Kobe, Hyogo 658-8501, Japan}

\author{Hiromichi Okon}
\affiliation{Center for Astrophysics | Harvard \& Smithsonian, 60 Garden Street, Cambridge, MA 02138, USA}

\author{Robert Fisher}
\affiliation{Department of Physics, University of Massachusetts Dartmouth, 285 Old Westport Road, North Dartmouth, MA 02740, USA}
\affiliation{Institute for Theory and Computation, Harvard-Smithsonian Center for Astrophysics, 60 Garden Street, Cambridge, MA 02138, USA}
\affiliation{Kavli Institute for Theoretical Physics, Kohn Hall, University of California at Santa Barbara, Santa Barbara, CA 93106, USA}

\author{Robert Petre}
\affiliation{NASA Goddard Space Flight Center, Code 662, Greenbelt, MD 20771, USA}

\author{Brian J. Williams}
\affiliation{NASA Goddard Space Flight Center, Code 662, Greenbelt, MD 20771, USA}



\begin{abstract}
The supernova remnant (SNR) 3C\,397 is thought to originate from a Type Ia supernova (SN Ia) 
explosion of a near-Chandrasekhar-mass ($M_{\rm Ch}$) progenitor, based on the 
enhanced abundances of Mn and Ni revealed by previous X-ray study with \Suzaku. 
Here we report follow-up \XMM\ observations of this SNR, conducted with the aim 
of investigating the detailed spatial distribution of the Fe-peak elements. 
We have discovered an ejecta clump with extremely high abundances of Ti and Cr, 
in addition to Mn, Fe, and Ni, in the southern part of the SNR. 
The Fe mass of this ejecta clump is estimated to be $\sim$\,0.06\,$M_{\odot}$, 
under the assumption of a typical Fe yield for SNe Ia (i.e., $\sim$\,0.8\,$M_{\odot}$). 
The observed mass ratios among the Fe-peak elements and Ti 
require substantial neutronization that is achieved only in the innermost regions of 
a near-$M_{\rm Ch}$ SN Ia with a central density of $\rho_c \sim 5 \times 10^9$\,g\,cm$^{-3}$, 
significantly higher than typically assumed for standard near-$M_{\rm Ch}$ SNe Ia 
($\rho_c \sim 2 \times 10^9$\,g\,cm$^{-3}$). 
The overproduction of the neutron-rich isotopes (e.g., $^{50}$Ti and $^{54}$Cr) is significant 
in such high-$\rho_c$ SNe Ia, with respect to the solar composition.
Therefore, if 3C\,397 is a typical high-$\rho_c$ near-$M_{\rm Ch}$ SN Ia remnant, 
the solar abundances of these isotopes could be reproduced by the mixture
of the high- and low-$\rho_c$ near-$M_{\rm Ch}$ and sub-$M_{\rm Ch}$ Type Ia events, with $\lesssim$\,20\,\% being high-$\rho_c$ near-$M_{\rm Ch}$.

\end{abstract}



\section{Introduction} \label{sec:intro}

Thermonuclear explosions of carbon-oxygen white dwarfs (WDs), 
or Type Ia supernovae (SNe Ia), are important astrophysical phenomena, 
because of their roles as standardizable candles for cosmology \citep[e.g.,][]{Phillips1993} 
and major sources of Fe in the cosmic chemical enrichment \citep[e.g.,][]{Kobayashi2020}. 
Despite decades of intense effort, many properties of SNe Ia, including how their progenitors 
evolve and explode, remain unsolved \citep[e.g.,][]{Maeda_Terada2016}. 
Given the relative uniformity in their observed properties, SNe Ia were formerly thought to 
originate always from near-Chandrasekhar-mass ($M_{\rm Ch}$) WDs that had grown 
via mass accretion from their binary companion \citep[e.g.,][]{Nomoto1984}. 
However, recent theoretical and observational work has provided many pieces of evidence 
in favor of sub-$M_{\rm Ch}$ progenitors as the major contributors to SNe Ia
(e.g., \citealt{Ruiter2009, Badenes_Maoz2012}, but see \citealt{Hachisu2012} and 
\citealt{Nomoto_Leung2018} for counter arguments).
In particular, element composition of several dwarf galaxies requires sub-Mch be dominant \citep{McWilliam2018, Kirby2019}.

Nevertheless, explosions of near-$M_{\rm Ch}$ WDs are still necessary from 
the perspective of the chemical evolution in the Milky Way
\citep[e.g.,][]{Seitenzahl2013a, Kobayashi2020, Palla2021} 
and galaxy clusters \citep[e.g.,][]{Mernier2016,Hitomi2017}, and even evolved dwarf spheroidal galaxies \citep{de_los_Reyes2020}
This is because efficient production of neutron-rich isotopes of 
the Fe-peak elements, such as $^{54}$Cr, $^{55}$Mn, and $^{58}$Ni, 
require electron capture processes that take place only in the dense 
core of near-$M_{\rm Ch}$ SNe Ia \citep[e.g.,][]{Leung_Nomoto2018, Leung_Nomoto2020}, 
and the contributions of sub-$M_{\rm Ch}$ SNe Ia cannot account for 
the observed abundances of these species.
Evidence of these neutron-rich isotopes has been 
searched for in the late-time optical/infrared light curves of individual SN Ia events.
\citep[e.g.,][]{Dimitriadis2017, Maguire2018}.
It is also suggested that future X-ray observations of nearby SNe Ia will be 
capable of detecting Mn K$\alpha$ lines from the decay of $^{55}$Fe, 
providing a powerful diagnostic for distinguishing between near-$M_{\rm Ch}$ 
and sub-$M_{\rm Ch}$ explosions (\citealt{Seitenzahl2015}, but see also \citealt{Lach2020} and \citealt{Gronow2021}, 
who argued that solar or super-solar Mn/Fe ratios can be achieved by sub-$M_{\rm Ch}$ explosions as well).

Observations of supernova remnants (SNRs) offer a complementary approach for constraining 
SN Ia progenitor mass and density. Although their X-ray spectra of Type Ia SNR do not distinguish 
among isotopes of each element, mass ratios among the Fe-peak elements in the 
ejecta can be accurately measured, since the plasma is optically thin 
and physics of dominant atomic processes is relatively well understood.
In fact, \Suzaku\ observations of the SN Ia remnant 3C\,397 detected strong K-shell 
fluorescence from the Fe-peak elements, and revealed that the observed mass ratios 
of Mn/Fe and Ni/Fe require substantial contributions from the electron capture elements, thereby ruling out the sub-$M_{\rm Ch}$ scenario for this specific SNR \citep{Yamaguchi2015}.

Following the observational result with \Suzaku, \cite{Dave2017} and \cite{Leung_Nomoto2018} 
theoretically investigated the effect of the WD central density ($\rho_{\rm c}$) 
on the resulting nucleosynthesis yields, and found that the mass ratios among the Fe-peak elements observed in 3C\,397 can be reproduced by an explosion 
of a near-$M_{\rm Ch}$ WD with a central density of $\rho_{\rm c} \approx 5 \times 10^9$\,g\,cm$^{-3}$, significantly higher 
than usually assumed in models of `standard' near-$M_{\rm Ch}$ SNe Ia 
($\sim 2 \times 10^9$\,g\,cm$^{-3}$: 
e.g., \citealt{Iwamoto1999}; \citealt{Bravo_Martinez2012}).
It is also notable that such high-density SN Ia models predict efficient 
production of $^{50}$Ti and $^{54}$Cr, in addition to $^{55}$Mn and $^{58}$Ni, 
due to the extremely high neutronization realized at the innermost region of 
the exploding WD \citep{Dave2017, Leung_Nomoto2018}. 
Therefore, detection of ejecta with enhanced abundances of Ti and Cr 
would provide further evidence of the high-density near-$M_{\rm Ch}$ progenitor, 
and place a more stringent constraint on the central density \citep[e.g.,][]{Sato2020}.

In this Letter, we present \XMM\ observations of 3C\,397 conducted to use 
its superior angular resolution to \Suzaku\ to search for highly-neutronized ejecta.
This SNR is located near the Galactic plane at a distance of $\sim$\,8\,kpc 
\citep{Leahy_Ranasinghe2016} and is well established as a middle-aged SN Ia remnant 
\citep{Yamaguchi2014a, Martinez2020}. 
Its angular size is about $5' \times 3'$.
Because of the aim of this work, we focus exclusively on the imaging spectroscopic analysis 
of the K-shell emission from Ti and the Fe-peak elements in some small regions. 
The uncertainties quoted in the text and table and the error bars in the figures 
represent the 1\,$\sigma$ confidence level.

\section{Observation and Data Reduction} \label{sec:obs}

We conducted a deep observation of 3C\,397 using the European Photon Imaging Camera (EPIC) on board \XMM\ in October 2018 (Obs.ID: 0830450101).
The EPIC consists of three charge-coupled devices, the MOS1, MOS2 \citep{Turner2001}, and pn \citep{Struder2001}.
The angular resolution of the telescopes is $\sim$\,$15''$ (half-power diameter), 
good enough to resolve small spatial structures of 3C\,397. 

We reprocessed all the data using the {\tt emchain} and {\tt epchain} tasks 
(for the MOS and pn, respectively) 
in the version 18.0.0 of the \XMM\ Science Analysis Software with the Current Calibration Files.
Event selection was conducted based on the standard screening criteria.
To eliminate data with high background rates, we filtered out observation periods 
when the count rate in the 0.2--12.0\,keV is higher than 8\,cnt\,s$^{-1}$ (MOS1), 
10\,cnt\,s$^{-1}$ (MOS2), and 80\,cnt\,s$^{-1}$ (pn).
The resulting effective exposure is $\sim$\,130\,ks, more than 90\,\% of the unfiltered exposure, for all the instruments.


\section{Analysis and Results} \label{sec:res}

\begin{figure*}[htbp]
 \centering
 \includegraphics[width=17.0cm]{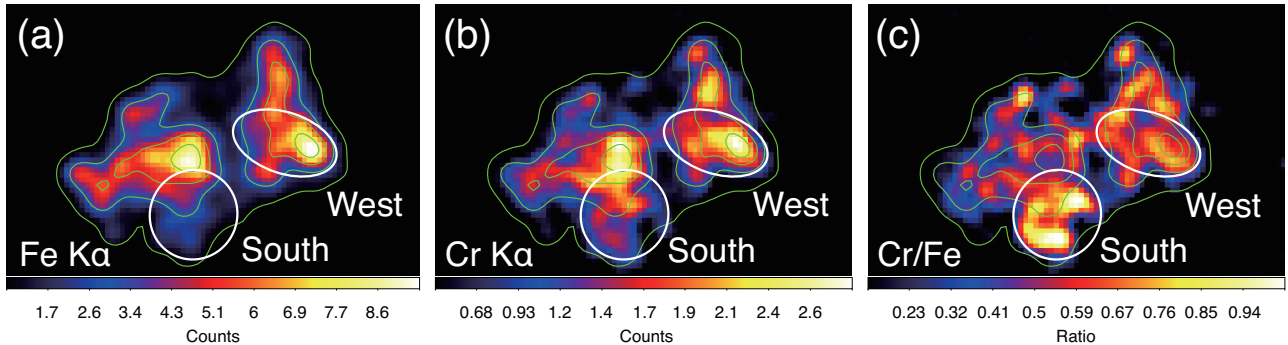}
 \caption{{\it XMM-Newton}/EPIC MOS images of the SNR 3C\,397 in the (a) Fe K$\alpha$ (6.4--6.7\,keV) and (b) Cr K$\alpha$ (5.4--5.7\,keV) bands. \  (c) Spatial distribution of the flux ratio between the Cr K$\alpha$ (5.4--5.7\,keV) and Fe K$\alpha$ (6.4--6.7\,keV) emission. The overplotted contours in (b) and (c) are the Fe K$\alpha$ flux image (same as panel (a)). The white ellipses indicate where the spectra shown in Figure~2 are extracted.}
 \label{fig1}
\end{figure*}

Figure~1(a) shows an EPIC/MOS image of 3C\,397 in the 6.4--6.7\,keV band, 
corresponding to the Fe K$\alpha$ fluorescence from intermediately ionized Fe. There are two bright regions at 
the west and center of the SNR, as reported in previous work
\citep{Chen1999, Safi2005}. 
We then investigate the spatial distribution of the other Fe-peak elements, 
and find that the morphology of the Cr K$\alpha$ emission is slightly different 
from that of the Fe K$\alpha$ emission (i.e., Figure~1(a)). 
To highlight this difference, we generate in Figure~1(b) a Cr-to-Fe flux ratio map 
by dividing the 5.4--5.7-keV image by the 6.4--6.7-keV image. 
The highest ratio is found in the south region. 

\begin{figure}[htbp]
 \centering
 \includegraphics[width=8.0cm]{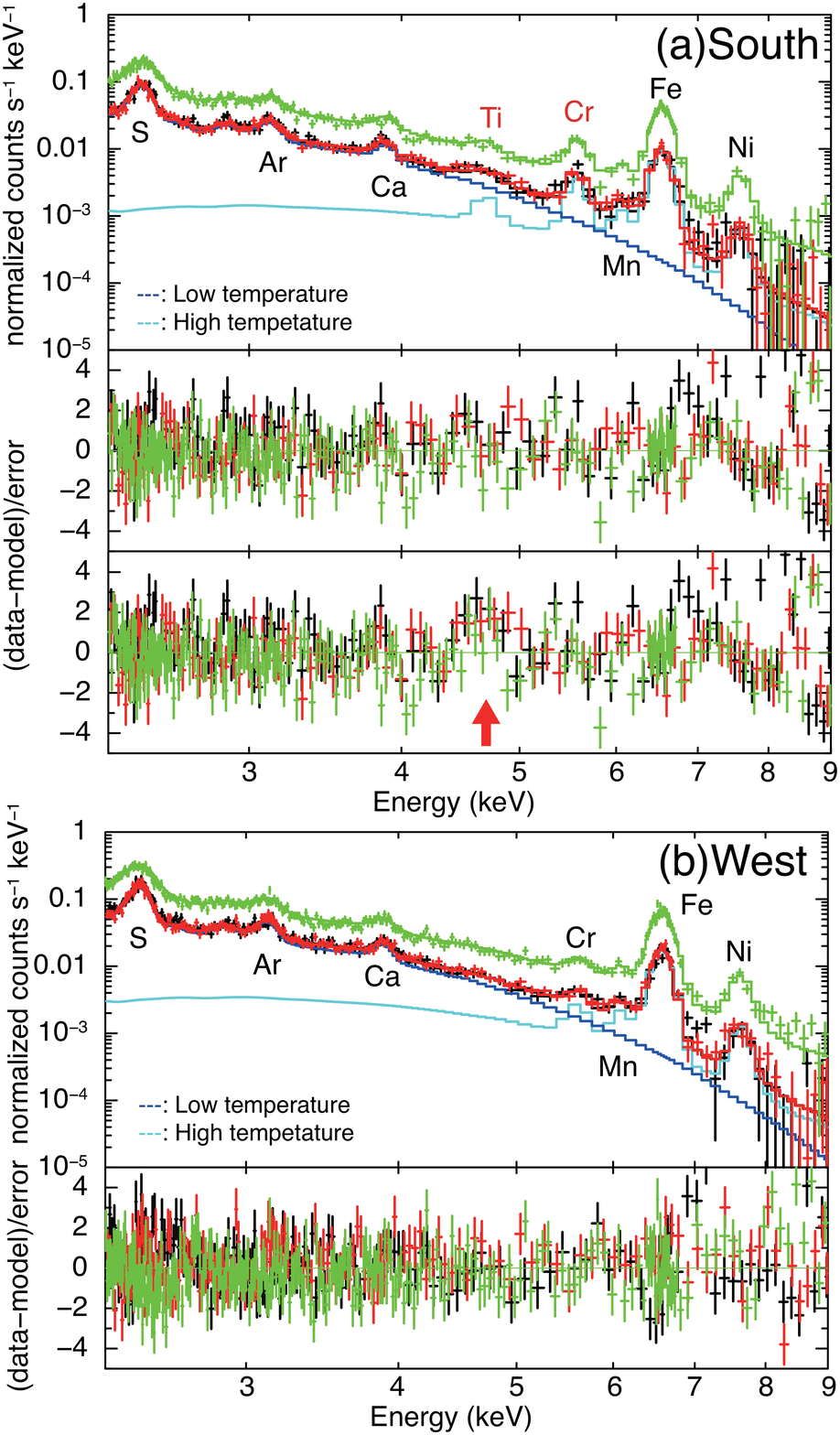}
 \label{fig:spec}
 \caption{EPIC spectra in the 2.3--9.0\,keV band extracted from the (a) South and (b) West regions given in Figure~1. Black, red, and green are the MOS1, MOS2, and pn, respectively. The best-fit models for the low- and high-temperature components are given as blue and cyan lines, respectively. The bottom window of panel (a) shows the residuals from the best-fit model without Ti K$\alpha$ emission, where the significant excess in the data is found around 4.7\,keV (red arrow).}
 \label{spec}
\end{figure}

For more quantitative study, we extract EPIC spectra 
from the two representative regions, ``South'' and ``West'' in Figure~1.
The former is where the highest Cr/Fe ratio is found, and the latter is 
where the surface brightness of both Fe K$\alpha$ and Cr K$\alpha$
emission is highest.
Figures~2(a) and 2(b) show the background-subtracted spectra of the South and West, 
respectively. Since our primary goal is to determine the mass ratios among the Fe-peak 
elements and Ti, the following spectral analysis focuses on the 2.3--9\,keV band 
so that we can model their K-shell emission as well as the underlying continuum emission adequately. 
The background data are taken from the outside of the SNR; the choice of 
background regions does not affect the following results significantly. 
In both spectra, K$\alpha$ lines of Cr, Mn, Fe, and Ni are clearly resolved. 
We confirm that the Cr K$\alpha$ emission is much stronger in the South than 
in the West, consistent with Figure~1(b). 
We also detect an emission feature around 4.7\,keV in the South spectra, 
likely originating from the K$\alpha$ fluorescence of stable Ti.

The spectra of both regions are fitted with a model of optically thin thermal plasma in 
non-equilibrium ionization (NEI), using the XSPEC software version 12.11.00 \citep{Arnaud1996}. 
The fitting is performed based on the $C$-statistic \citep{Cash1979} on unbinned spectra, 
although the spectra shown in Figure~2 are binned for clarity. 
For the foreground absorption, we introduce the Tuebingen-Boulder interstellar medium absorption ({\tt TBabs}) model \citep{Wilms2000}
with the hydrogen column density (\NH) fixed at $3 \times 10^{22}$\,cm$^{-2}$ 
\citep{Safi2005}. 
We start fitting the spectra with a single NEI component, with free parameters of 
electron temperature ($kT_{\rm e}$), ionization timescale ($n_{\rm e}t$), normalization, 
and abundances of S, Ar, Ca, Ti, Cr, Mn, Fe, and Ni. This model fits the data well below 4\,keV, 
but fails to reproduce the centroid of the emission lines of Ti and the Fe-peak elements or the continuum flux in $>$ 4\,keV. 
We thus introduce another NEI component with different electron temperature and 
ionization timescale. The abundances of the heavy elements are also left independent 
between the two components. This addition significantly improves the c-stat values 
from 4406.02 to 4151.96 for the South and from 4475.16 to 4291.93 for the West.
The results obtained for the two regions are summarized as follows:
The low-$kT_{\rm e}$ ($\sim$\,1\,keV), high-$n_{\rm e}t$ ($\sim$\,$10^{11}$\,cm$^{-3}$\,s) 
component reproduces the emission lines of S, Ar, and Ca, but does not significantly contribute 
to the emission of Ti, Cr, Mn, Fe, and Ni with their abundances of consistent with zero. 
The high-$kT_{\rm e}$ ($\sim$\,5\,keV), low-$n_{\rm e}t$ ($\sim$\,$3 \times 10^{10}$\,cm$^{-3}$\,s) 
component, on the other hand, yields extremely high ($\gtrsim$\,1000\,solar) abundances of 
Ti and the Fe-peak elements, implying that this component is dominated by these elements with 
no admixture of the lighter elements including H and He.
We thus fix the abundances of Ti and the Fe-peak elements in the low-$kT_{\rm e}$ component 
to zero. For the high-$kT_{\rm e}$ component, we fix its Fe abundance to $1 \times 10^8$~solar 
(which mimics thermal emission from a pure-metal plasma) and allow the abundances of 
Ti, Cr, Mn, and Ni to vary freely, obtaining the best-fit results given in Table~1.

Finally, we convert the observed abundances to the mass ratios among the elements as follows:
\[
    \frac{M_{i}}{M_{\rm Fe}} = \frac{m_{i}}{m_{\rm Fe}} \, \frac{A_{i}}{A_{\rm Fe}}\left(\frac{n_{i}}{n_{\rm Fe}}\right)_{\odot} (i = \rm Ti,\,Cr,\,Mn,\,Ni),
\]
where $m_i$ is the atomic weight, $A_{\rm i}$ is the observed abundance (Table 1), 
and $(n_{i}/n_{\rm Fe})_{\odot}$ is the elemental number ratio in the solar abundance
of \cite{Anders_Grevesse1989}. The resulting mass ratios are also given in Table~1. 
In this conversion, we have assumed the average atomic weight in the solar system 
(i.e., $m_{\rm Ti} = 47.9$, $m_{\rm Cr} = 52.0$, $m_{\rm Mn} = 54.9$, 
$m_{\rm Fe} = 55.8$, and $m_{\rm Ni} = 58.7$). 
Although the isotope ratios expected for SN Ia yields are not identical to 
the solar composition (see \S4), the difference between them is negligibly small 
(e.g.: $\lesssim$ 4\% for Ti) compared to the statistical uncertainty.

\begin{table}[htpb]
{
 \centering
  \begin{tabularx}{8.4cm}{ccc} \hline \hline
    Region & South & West \\ \hline
    
    \multicolumn{3}{c}{Low-temperature component} \\ \hline
    
    $kT_{\rm e}\ [\rm keV]$ & 
    $1.14_{-0.18}^{+0.06}$ & 
    $1.30_{-0.06}^{+0.03}$ \\
    
    $n_{\rm e}t\ [10^{11}\, \rm cm^{-3}\,s]$ & 
    $2.2_{-0.5}^{+0.5}$ & 
    $1.2_{-0.1}^{+0.2}$ \\
    
    $n_{\rm e}n_{\rm H}V^{\rm a}\ [10^{57}\, \rm cm^{-3}]$ &
    $4.9_{-0.3}^{+0.9}$ & 
    $7.0_{-0.3}^{+0.3}$ \\
    
    S\ [solar] & 
    $1.15_{-0.06}^{+0.06}$ & 
    $1.19_{-0.05}^{+0.07}$ \\
    
    Ar\ [solar] & 
    $0.89_{-0.10}^{+0.13}$ & 
    $1.03_{-0.07}^{+0.09}$ \\
    
    Ca\ [solar] & 
    $1.55_{-0.19}^{+0.20}$ & 
    $1.83_{-0.18}^{+0.22}$ \\ \hline
    
    \multicolumn{3}{c}{High-temperature component} \\ \hline
    
    $kT_{\rm e}\ [\rm keV]$ & 
    $4.6_{-1.8}^{+1.4}$ & 
    $3.4_{-0.6}^{+0.3}$ \\
    
    $n_{\rm e}t\ [10^{10}\, \rm cm^{-3}\,s]$ & 
    $2.9_{-0.2}^{+0.7}$ & 
    $5.1_{-0.1}^{+0.7}$ \\
    
    $n_{\rm e}n_{\rm H}V^{\rm a}\ [10^{49}\, \rm cm^{-3}]$ & 
    $4.0_{-1.1}^{+4.6}$ & 
    $8.5_{-0.9}^{+3.0}$ \\
    
    $\rm{Velocity^{b}} [\rm 10^{3}\ km\,s^{-1}]$ & 
    $2.8_{-0.3}^{+0.3}$ & 
    $2.7_{-0.2}^{+0.2}$ \\
    
    Ti\ [$10^{8}$solar] & 
    $8.0_{-2.0}^{+2.2}$ & 
    $<0.9$ \\
    
    Cr\ [$10^{8}$solar] & 
    $11.4_{-0.1}^{+0.1}$ & 
    $3.6_{-0.4}^{+0.5}$ \\
    
    Mn\ [$10^{8}$solar] & 
    $9.8_{-1.7}^{+1.7}$ & 
    $9.7_{-1.2}^{+1.2}$ \\
    
    Fe\ [$10^{8}$solar] & 
    $1.0$\ (fixed) & 
    $1.0$\ (fixed) \\
    
    Ni\ [$10^{8}$solar] & 
    $4.5_{-0.7}^{+1.7}$ & 
    $6.8_{-0.7}^{+1.3}$ \\ \hline
    
    \multicolumn{3}{c}{Statistic} \\ \hline
    cstat & 4151.96 & 4291.93 \\ 
    $d.o.f.$ & 4001 & 4002 \\ \hline
    
    \multicolumn{3}{c}{Mass ratio} \\ \hline
    Ti/Fe & 
    $0.014_{-0.004}^{+0.004}$ & 
    $<0.002$ \\
    
    Cr/Fe & 
    $0.106_{-0.009}^{+0.011}$ & 
    $0.034_{-0.004}^{+0.004}$ \\
    
    Mn/Fe & 
    $0.051_{-0.009}^{+0.009}$ & 
    $0.050_{-0.006}^{+0.006}$ \\
    
    Ni/Fe & 
    $0.18_{-0.03}^{+0.07}$ & 
    $0.27_{-0.03}^{+0.05}$ \\ \hline 
    
  \end{tabularx}
  }
  \caption{The best-fit parameters from model fitting and derived mass ratios.}
  \tablecomments{$^{\rm a}$The volume emission measure, where $n_{\rm e}$ is the electron density, $n_{\rm H}$ the hydrogen density, and $V$ is the volume of the emission region. $^{\rm b}$Velocity dispersion in the line of sight for the Doppler broadening of the emission lines.}
  \label{table:fitpara}
\end{table}

\section{Interpretation and Discussion} \label{sec:dis}

Thanks to the excellent angular resolution and sensitivity of the \XMM/EPIC, we have 
succeeded in spatially resolving an ejecta clump with an enhanced Cr abundance at the South 
of the SNR 3C\,397. Its Cr/Fe mass ratio is about three times higher than that in the West 
as well as the previous \Suzaku\ measurement for the entire SNR \citep{Yamaguchi2015}. 
The spectra also suggest that this ejecta clump has an extremely high abundance of stable Ti.
To our knowledge, this is the first detection of stable Ti from this SNR and the second after 
Tycho's SNR \citep{Miceli2015} among Type Ia SNRs\footnote{Emission from stable Ti has recently been detected in a core-collapse SNR, Cassiopeia A \citep{Sato2021}.
This emission is thought to originate predominantly from $^{48}$Ti (a major product of $\alpha$-rich freeze out), whereas the stable Ti detected in 3C\,397 is likely dominated by $^{50}$Ti.}.
Note however that the detection in Tycho's SNR has been disputed by \cite{Yamaguchi2017} 
and \cite{Sato2020}. This is because the centroid energy of the detected line ($\sim$\,4.9\,keV), 
which corresponds to the Ly$\alpha$ emission of H-like Ti, 
is unnaturally higher than expected for the plasma condition in Tycho's SNR.

The Fe K$\alpha$ flux ($F$) of the ejecta clump is estimated to be 
$1.7 \times 10^{-13}$\,erg\,cm$^{-2}$\,s$^{-1}$, 
which is 13\,\% of the Fe K$\alpha$ flux from the entire SNR 
($1.3 \times 10^{-12}$\,erg\,cm$^{-2}$\,s$^{-1}$). 
The observed solid angle $(\Omega)$ of the ejecta clump is $1.5 \times 10^{-7}$\,sr, 
about 12\,\% of the solid angle subtended by the entire SNR. 
For a pure metal plasma, the relationship between the Fe mass ($M_{\rm Fe}$) and $F$ 
is given as $M_{\rm Fe} \propto (F\cdot V)^{1/2} \sim F^{1/2}\cdot \Omega^{3/4}$ 
(where $V$ is the plasma volume, within which uniform density distribution is assumed). 
Therefore, we estimate the Fe mass in the ejecta clump to be 
$\sim$\,7\% of the total Fe ejecta mass, corresponding to $\sim$\,0.06 $M_{\odot}$, 
under an assumption of a typical SN Ia Fe yield (i.e., $\sim$\,0.8\,$M_{\odot}$).

Our spectral analysis indicates that Ti and all the Fe-peak elements compose a single (high-$kT_{\rm e}$) plasma component, suggesting that these elements 
are generated by a common nucleosynthesis regime. In the case of near-$M_{\rm Ch}$ 
SNe Ia, Ti and Cr (or their parent nuclei) are synthesized in either incomplete 
Si burning or neutron-rich nuclear statistical equilibrium (n-rich NSE). 
The former takes place in the relatively low temperature ($T \sim 5 \times 10^9$\,K) and 
low density ($\rho \sim 10^7$\,g\,cm$^{-3}$) environment and produces $^{48}$Cr and $^{52}$Fe 
via $\alpha$-processes, which eventually decay into $^{48}$Ti and $^{52}$Cr, respectively. 
However, production of stable Ni requires higher temperature and density 
\citep[e.g,][]{Iwamoto1999}.
The n-rich NSE, on the other hand, takes place in a high temperature 
($T \gtrsim 5.5 \times 10^9$\,K) and high density 
($\rho \gtrsim 3 \times 10^8$\,g\,cm$^{-3}$) environment and directly produces 
the neutron-rich isotopes of all the elements found in the high-$kT_{\rm e}$ component (e.g, $^{50}$Ti, $^{52}$Cr, $^{54}$Cr, $^{55}$Mn, $^{54}$Fe, $^{56}$Fe, $^{58}$Ni). 
Therefore, we conclude that the ejecta clump in the 3C\,397 South is an n-rich NSE product 
originating from the dense core of the near-$M_{\rm Ch}$ progenitor.

It is theoretically suggested that efficient production of stable Ti and Cr 
in SNe Ia is achieved only when the WD central density is sufficiently high  
($\rho_c \gtrsim 3 \times 10^9$\,g\,cm$^{-3}$). 
This is illustrated in Figure~3, numerical calculations of SN Ia nucleosynthesis 
with different $\rho_c$ assumed \citep{Leung_Nomoto2018}. These models are based on 
two-dimensional hydrodynamical calculations of the delayed-detonation explosion \citep[e.g.,][]{Khokhlov1991} of a near-$M_{\rm Ch}$ WD with the solar metallicity. 
Note that, in the n-rich NSE, the metallicity effect on the resulting nucleosynthesis yields is negligible \citep[e.g.,][]{Yamaguchi2015}.
The models assume central ignition of the carbon deflagration and a detonation transition 
density of $\sim 2 \times 10^7$\,g\,cm$^{-3}$.
Figures~3(a) and 3(b) show Ti/Fe mass ratios plotted against the density when each tracer particle 
achieves the maximum temperature during the nuclear burning, ($\rho_{T_{\rm Max}}$), 
for the cases of $\rho_c = 1 \times 10^9$\,g\,cm$^{-3}$ (low density) 
and $\rho_c = 5 \times 10^9$\,g\,cm$^{-3}$ (high density), respectively. 
The mass ratio observed in the 3C\,397 South region is also indicated with the red area. 
Similar comparisons for the Cr/Fe mass ratios are given in Figures~3(c) and 3(d) 
for the low and high density cases, respectively.
It is confirmed that only the innermost ejecta in the high density case achieve the observed
mass ratios of $M_{\rm Ti}/M_{\rm Fe} \sim 0.01$ and $M_{\rm Cr}/M_{\rm Fe} \sim 0.1$. 

At lower densities around $1 \times 10^9$\,g\,cm$^{-3}$, the neutronized nucleosynthesis products are dominated by 
$^{58}$Ni, since the neutron excess in the NSE is not high enough to produce stable Ti and Cr.
For this reason, the mass ratios of Ti/Ni and Cr/Ni at the innermost part offer the most sensitive indicators of the central density of near-$M_{Ch}$ WDs \citep[e.g.,][]{Leung_Nomoto2018, Mori2018}.
\begin{figure*}[htpb]
 \centering
 \includegraphics[width=17cm]{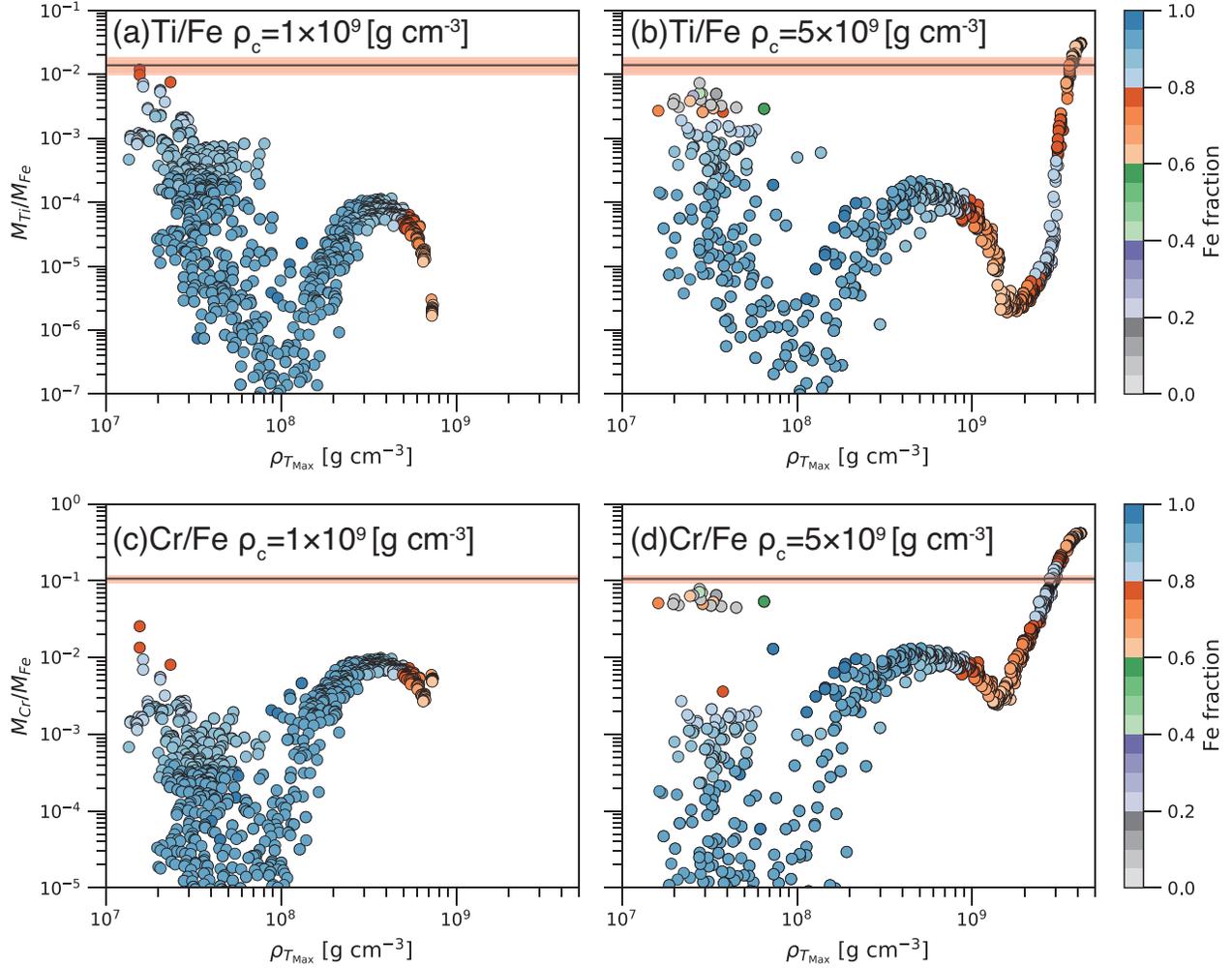}
 \caption{{\it Top:} Relationship between the Ti/Fe mass ratio and $\rho_{T_{\rm Max}}$ (the density when each tracer particle achieves the maximum temperature) predicted by the delayed-detonation near-$M_{\rm Ch}$ SN Ia models with (a) $\rho_c = 1 \times 10^9$\,g\,cm$^{-3}$ and (b) $\rho_c = 5 \times 10^9$\,g\,cm$^{-3}$ \citep{Leung_Nomoto2018}. Only the particles that satisfy $T_{\rm Max} \geq 5.5 \times 10^9$\,K are plotted. The Fe fraction in each particle is indicated by color. The observed mass ratio is indicated with the horizontal lines (best-fit) and red regions (1$\sigma$ statistical uncertainty). 
 {\it Bottom:} Same as the top panels, but for the Cr/Fe mass ratio. Panels (c) and (d) are for $\rho_c = 1 \times 10^9$\,g\,cm$^{-3}$ and $\rho_c = 5 \times 10^9$\,g\,cm$^{-3}$, respectively.}
 \label{fig:model}
\end{figure*}
In Figure~4, we plot these mass ratios (and $M_{\rm Mn}/M_{\rm Ni}$) as a function of 
$\rho_c$, predicted for the innermost ejecta by the models of \cite{Leung_Nomoto2018}.
Here we define the ``innermost ejecta'' as the ejecta in the highest $\rho_{T_{\rm Max}}$ regions whose integrated Fe mass is 7\,\% of the total Fe yield (or $M_{\rm Fe} \sim 0.06\,M_{\odot}$), 
so that we can directly compare the theoretical predictions with the observed mass ratios for 3C 397 South ejecta clump. 
We find that all the ratios require a central density of $> 3 \times 10^{9}$\,g\,cm$^{-3}$. 
In particular, the Ti/Fe ratio indicates $\rho_c \approx 5 \times 10^{9}$\,g\,cm$^{-3}$, 
suggesting strongly the high-density scenario for the progenitor of 3C 397.
We note that the mass ratios given in Figure~4 do not point towards a single value of $\rho_{c}$. However, in addition to the white dwarf
central density, the nucleosynthetic abundances depend on a larger parameter space, including the number of ignition kernels and the detonation transition density among others. 
Also, in the accreting WD models of SNe Ia, $\rho_c$ depends
on the accretion rate, the initial mass of the WD, and the angular
momentum of the WD \citep{Nomoto1982, Benvenuto2015}.  It would be
important to clarify the evolutionary origin of such high $\rho_c$
near-$M_{\rm Ch}$ SNe Ia as 3C397 in future study.

\cite{Martinez2017} argued that the predicted mass ratio of 
Ca/S was sensitive to the progenitor metallicity and claimed a supersolar metallicity for the progenitor of 3C\,397.
We have also investigated the nucleosynthesis calculations of \cite{Dave2017} 
and \cite{Leung_Nomoto2018}, and confirmed that, regardless of central density, 
the predicted Ca/S mass ratios for the entire ejecta show a metallicity dependence 
similar to that reported in \cite{Martinez2017}. 
In contrast, the yields of the neutron-rich Fe peak elements (i.e., Ti and Cr) in the innermost ejecta are sensitive only to the central density \citep{Leung_Nomoto2018}. 
Therefore, in principle, both progenitor metallicity and central density can be 
independently constrained using the mass ratios of these diagnostic elements 
measured based on spatially resolved spectroscopy. 

\begin{figure}[htbp]
 \centering
 \includegraphics[width=8.0cm]{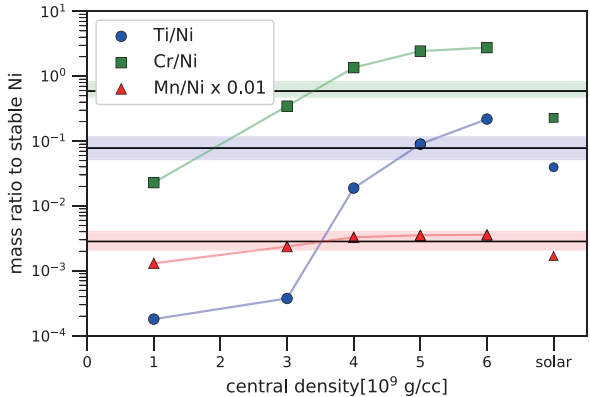}
 \caption{Predicted mass ratios of Ti/Ni (blue), Cr/Ni (green), and Mn/Ni (red) in the innermost ejecta (see text) as a function of the WD central density. The observed mass ratios are indicated with the horizontal lines and colored areas. 
 The solar values of \cite{Lodders2003} are also plotted at the rightmost for comparison.
}
 \label{fig:ratio}
\end{figure}

Interestingly, the neutronized ejecta clump is located at the southern outermost part
of the SNR (see Figure~1), despite its plausible origin of the densest WD core. 
This is, in fact, qualitatively consistent with predictions of multi-dimensional 
hydrodynamical calculations for an asymmetric SN Ia that involve delayed detonation 
\citep[e.g.,][]{Maeda2010, Seitenzahl2013b}. In these models, 
the nuclear burning that takes place during the initial deflagration phase makes 
the burnt materials hotter and less dense than the surrounding unburnt materials, 
leading to the buoyant rise of the deflagration ash (i.e., n-rich NSE products) to 
the WD surface.

As mentioned in \S\ref{sec:intro}, the nucleosynthesis that occurs in sub-$M_{\rm Ch}$ Type Ia SNe does not produce the observed abundances of Fe-peak isotopes; a contribution from near-$M_{\rm Ch}$ explosions is needed. \citep[e.g.,][]{Seitenzahl2013a, Kobayashi2020, Hitomi2017}.
In taking into account nucleosynthesis of near-$M_{\rm Ch}$ SNe Ia,
we should note that the abundance ratios of Fe-peak elements depend
sensitively on $\rho_c$ of the WD. For $\rho_c \sim 2 \times 10^9$\,g\,cm$^{-3}$
(low-$\rho_c$), the abundance pattern is close to the solar, while for
$\rho_c \approx 5 \times 10^9$\,g\,cm$^{-3}$ (high-$\rho_c$), the
neutron-rich isotopes are overproduced with respect to the solar
ratios.
Suppose that there are three types of SNe Ia, i.e., (1) high-$\rho_c$
near-$M_{\rm Ch}$ SNe Ia, (2) low-$\rho_c$ near-$M_{\rm Ch}$ SNe Ia, and (3) sub-$M_{\rm Ch}$ SNe Ia. Then the ratios between these
three types are subject to the nucleosynthesis yields of each SN type. 
If 3C\,397 is the remnant of a typical high-$\rho_c$ near-$M_{\rm Ch}$ SN Ia, we can
obtain more quantitative constraints on the frequency of such SN Ia by applying the nucleosynthesis yields of
$\rho_c = 5 \times 10^{9}$\,g\,cm$^{-3}$ model \citep{Leung_Nomoto2018}.
This model overproduces the neutron-rich isotopes with the ratios of
$^{50}$Ti/$^{56}$Fe and $^{54}$Cr/$^{56}$Fe being 5.3 and 18.6 times the solar ratio, respectively \citep{Leung_Nomoto2018}.  On the other hand,
these ratios are negligibly small in the low-$\rho_c$ near-$M_{\rm Ch}$ SN Ia and
sub-$M_{\rm Ch}$ SN Ia, so that these types of SNe Ia contribute to
dilution of neutron-rich isotopes by producing $\sim 0.8 M_\odot$ Fe. Thus, the solar abundances of these isotopes are reproduced
by the mixture of the high-$\rho_c$ near-$M_{\rm Ch}$ SNe Ia for
$\lesssim$\,20\,\%, and the other SNe Ia for the rest.  To obtain further constraints
on the ratios among three types of SNe Ia needs to consider additional
elements.

It is worth noting that a departure from the solar composition in the neutron-rich 
isotopic abundances (e.g., $^{48}$Ca, $^{50}$Ti, $^{54}$Cr) is commonly seen in 
carbonaceous chondrites meteorites 
\citep[e.g.,][]{Dauphas2010, Dauphas2014, Warren2011, Nittler2018}.
These meteorites are thought to be composed of presolar grains that were injected 
by nearby supernovae to the protosolar disk. 
The literture argued that the isotopic anomalies observed in the carbonaceous chondrites 
could be explained by either high-density ($\rho_c \gtrsim 5 \times 10^9$\,g\,cm$^{-3}$)
SNe Ia or electron capture supernovae \citep[ECSNe][]{Nomoto_Leung2017}, but no determination has been made.
Our result potentially helps distinguish between the two scenarios, since it offers evidence for the existence of high-density SNe Ia.
It is also notable that \cite{Nittler2018} reported an extremely high abundance 
of $^{54}$Cr in the smallest ($\lesssim 80$\,nm) grains. 
Theoretically, such small grains tend to be produced by SNe Ia 
\citep{Nozawa2011}, qualitatively supporing the SNe Ia scenario for 
the origin of the isotopic anomalies.

\section{Conclusions} \label{sec:con}
We have presented an imaging spectroscopic study of the Fe-peak elements in 
the SNR 3C\,397, using a the deep \XMM\ observation conducted in 2018. 
An enhanced abundance of Cr is found in an ejecta clump in the southern part of 
the SNR, where we have also detected K-shell emission from stable Ti. 
The mass ratios of Ti/Fe, Cr/Fe, Mn/Fe, and Ni/Fe are obtained to be 
$0.014_{-0.004}^{+0.004}$, $0.106_{-0.009}^{+0.011}$, $0.050_{-0.009}^{+0.009}$, 
and $0.18_{-0.03}^{+0.07}$, respectively. 
The creation of elements with these abundances requires substantial neutronization that can be achieved only in the innermost regions of a near-$M_{\rm Ch}$ SN Ia with a central density of 
4--5\,$\times 10^9$\,g\,cm$^{-3}$. 
The ejecta clump location implies that the n-rich NSE products 
buoyantly rose to the stellar surface during the initial deflagration phase.
If a 3C\,397-like SN Ia is typical of high-$\rho_c$ near-$M_{\rm Ch}$ explosion,
the solar abundance ratios of $^{50}$Ti/$^{56}$Fe and $^{54}$Cr/$^{56}$Fe can
be reasonably explained by a combination of the high-$\rho_c$ near-$M_{\rm Ch}$, and the low-$\rho_c$ near-$M_{\rm Ch}$ and sub-$M_{\rm Ch}$ Type Ia events, with $\lesssim$\,20\,\% being high-$\rho_c$ near-$M_{\rm Ch}$.
Our results may also help identify the origin of isotopic anomalies observed in 
presolar grains that compose carbonaceous chondrites.


\acknowledgments
s
We are grateful to Ryota Fukai and Takaya Nozawa for discussion about isotopic compositions of meteorites and dust formation in SNe.
This work is supported by Grants-in-Aid for Scientific Research (KAKENHI) of 
the Japanese Society for the Promotion of Science (JSPS) grant Nos.\ 
JP19H00704 (H.Y.), JP20H00175 (H.Y.) JP19H01936 (T.T.), JP17K05382(K.N.) and JP20K04024 (K.N.).
S.C.L. acknowledges support from grants HST-AR-15021.001-A and 80NSSC18K1017.
K.N. and S.C.L. have been supported by the World Premier International Research Center
Initiative (WPI Initiative), MEXT, Japan.  
R.F. acknowledges support from NASA grant number 80NSSC18K1013. This work used the Extreme Science and Engineering Discovery Environment (XSEDE) Stampede 2 supercomputer at the University of Texas at Austin’s Texas Advanced Computing Center through allocation TG-AST100038, supported by National Science Foundation grant number ACI-1548562.




\end{document}